\documentclass[lettersize,journal]{IEEEtran}
\usepackage{amsmath,amsfonts,amssymb}
\usepackage{graphicx}
\usepackage{algorithmicx}
\usepackage{algorithm}
\usepackage{algpseudocode}
\usepackage{textcomp}
\usepackage{makecell}
\usepackage{multirow,multicol}
\usepackage[colorlinks]{hyperref}
\hypersetup{citecolor=blue}
\usepackage[table]{xcolor}
\usepackage{threeparttable}
\usepackage{balance}
\usepackage{booktabs}
\usepackage[mathscr]{eucal}
\begin{document}

\title{Finite-blocklength Fluid Antenna Systems With\\ Spatial Block-Correlation Channel Model}

\author{Zhentian Zhang,~\IEEEmembership{Graduate Student Member,~IEEE}, Kai-Kit Wong,~\IEEEmembership{Fellow,~IEEE}, David Morales-Jimenez,~\\\IEEEmembership{Senior Member,~IEEE}, Hao Jiang,~\IEEEmembership{Senior Member,~IEEE}, Pablo Ramírez-Espinosa,~\IEEEmembership{Member,~IEEE}, \\Chan-Byoung Chae,~\IEEEmembership{Fellow,~IEEE}, and Christos Masouros,~\IEEEmembership{Fellow,~IEEE}
\thanks{}
\thanks{Z. Zhang and H. Jiang are with the National Mobile Communications Research Laboratory, Southeast University, Nanjing, 210096, China and H.~Jiang is also with the School of Artificial Intelligence, Nanjing University of Information Science and Technology, Nanjing 210044, China. (e-mail: zhentianzhangzzt@gmail.com, jianghao@nuist.edu.cn).}
\thanks{K.-K. Wong and C. Masouros are with the Department of Electronic and Electrical Engineering, University College London, Torrington Place, WC1E 7JE, United Kingdom  (e-mails: \{kai-kit.wong, c.masouros\} @ucl.ac.uk). K.-K. Wong is also affiliated with the Yonsei Frontier Lab., Yonsei University, Seoul, 03722 South Korea.}
\thanks{D. Morales-Jimenez is with the Department of Signal Theory, Networking and Communications, University of Granada, Granada 18071, Spain (e-mail: dmorales@ugr.es).}
\thanks{P. Ramírez-Espinosa is with Telecommunications Research Institute (TELMA), University of Malaga, 29071 Malaga, Spain (e-mail: pre@ic.uma.es).}
\thanks{C.-B. Chae is with the School of Integrated Technology, Yonsei University, Seoul, 03722 South Korea (e-mail: cbchae@yonsei.ac.kr).}
\thanks{Corresponding authors: K.-K. Wong (kai-kit.wong@ucl.ac.uk)}}



\maketitle
\begin{abstract}
	\textcolor{blue}{Finite blocklength (FBL) effects are often overlooked in analyses based on outage probability, where the assumption of vanishing decoding error can lead to overly optimistic performance predictions. Although fluid antenna systems (FASs) have been shown to deliver substantial spatial diversity for various system designs, their behavior under FBL constraints has not been adequately studied. Leveraging the simplicity, tractability, and accuracy of the spatial block-correlation channel model for FAS, this work investigates the {\em random coding performance} limits of FBL-coded systems equipped with fluid antennas. We analytically characterize how FAS contributes to performance enhancement under the FBL assumption and validate the findings through numerical evaluations. Both analytical and simulation results reveal that FAS offers pronounced advantages over conventional systems employing multiple fixed antennas in FBL-coded communication scenarios.}
\end{abstract}


\section{Introduction}
\IEEEPARstart{F}{luid} antenna systems (FAS) represent a promising communication paradigm capable of achieving substantial spatial diversity by dynamically adjusting radiation characteristics through flexible antenna structures and adaptive protocol designs~\cite{FAS_tutorial}. Unlike traditional architectures that attempt to mitigate antenna correlation in compact spaces, FAS deliberately exploits such correlation to attain notable channel gains \cite{FAS_twc_21,BLER_expression}, offering enhanced service reliability with lower hardware complexity and simplified signaling procedures.
\par\indent \textcolor{blue}{Most existing studies focus on outage probability performance \cite{FAS_tutorial,Block-Correlation}, which is not applicable when finite blocklength (FBL) effects are present \cite{FBL1,FBL2}. Under the FBL regime, decoding errors do not vanish, and thus outage probability is not to be defined. Although \cite{FBL1} provides a closed-form block error rate (BLER) expression for random coding analysis, it requires extensive system-specific parameters, limiting its practical applicability.}
\par\indent \textcolor{blue}{In this work, we investigate the {\em random coding limits} of FAS-enabled systems under FBL assumptions by carefully analyzing and simplifying both the BLER expression and the channel-strength distribution. The main contributions are summarized as follows:}
\textcolor{blue}{\begin{itemize}
	\item[1)] We derive the exact PDF of the channel strength $|g_{\mathrm{FAS}}|^2$ in expression \eqref{eq:pdf_channel_response}, whose integral form is inherently complex. We also present a simplified numerical evaluation based on Gauss--Laguerre quadrature by \eqref{eq.GL_A}.
	\item[2)] To further reduce complexity, we introduce a Gamma Approximation for the block-correlation model in \eqref{eq.GA_pdf_cdf}, yielding a compact and computationally efficient PDF representation with high approximation accuracy.
	\item[3)] We derive a high-SNR approximation of the instantaneous BLER for short-packet/finite-length codes in \eqref{eq.BLER_approximation}, which eliminates the original integral form and substantially lowers computational overhead.
\end{itemize}}
\par\indent \textcolor{red}{{\em The reproducible simulation code used in this work can be accessed at:
		https://github.com/BrooklynSEUPHD/Finite-blocklength-Fluid-Antenna-Systems-With-Spatial-Block-Correlation-Channel-Model.git.} }
\par\indent The rest of the paper is organized as follows. Sec.~\ref{sec.2} introduces the block-correlation channel model for FBL-FAS. Sec.~\ref{sec.3} presents the main analytical results with detailed proofs. Numerical evaluations are provided in Sec.~\ref{sec.4}, and conclusions are drawn in Sec.~\ref{sec.5}.
\section{System Descriptions}\label{sec.2}
\textcolor{blue}{In this work, we consider a point-to-point transmission scenario (downlink or uplink) where a base station (BS) equipped with a conventional single antenna serves a single-antenna user equipment for exchange of $U$ bits of information assuming transmission with blocklength $M$. The UE employs a fluid antenna with $N$ preset configurable locations/{\em ports} that are evenly distributed along a linear array of length $W$, where $W$ denotes the antenna length normalized by the wavelength at the operating frequency. Each port is regarded as an ideal antenna; however, due to limited hardware resources such as the radio frequency (RF) chain, only one port will be activated at a time, i.e., a single RF chain is assumed.
\subsection{Finite-Blocklength Single Antenna FAS Signal Model}
The received signal can be expressed as
\begin{equation}
	\boldsymbol{y} = g_{u,k}\boldsymbol{x}_{u} + \boldsymbol{\eta},
\end{equation}
where $\boldsymbol{x}_{u}$ denotes the Gaussian codeword of the user with zero mean and variance $\sigma_c^2 = \tfrac{1}{M}$, such that $\|\boldsymbol{x}_{u}\|^2_2 = 1$. The term $\boldsymbol{\eta}$ represents the additive white Gaussian noise (AWGN) with zero mean and variance $\tfrac{\sigma^2_{\eta}}{M}$, i.e., $\|\boldsymbol{\eta}\|^2_2 = \sigma^2_{\eta}$. With a slight abuse of notation, $g_{u,k}$ denotes the $k$-th channel coefficient of the user, where $k \in \{1,2,\ldots,N\}$ represents the FAS port. The channel coefficients are Rayleigh distributed with variance $\sigma^2$ and correlated by correlation matrix $\boldsymbol{\Sigma}$. Moreover, the signal-to-noise ratio (SNR) is defined as:
\begin{equation}
	\mathrm{SNR} = \frac{\mathbb{E}\left\{\|g_{u,k}\boldsymbol{x}_{u}\|^2_2\right\}}{\mathbb{E}\left\{\|\boldsymbol{\eta}\|^2_2\right\}} 
	= \frac{\sigma^2}{\sigma^2_{\eta}}.
\end{equation}
\subsection{Spatial Block-Correlation Channel Model for Single Antenna $N$-Ports FAS}
 The spatial block-correlation channel model in \cite{Block-Correlation} is considered for the potential simplicity, modeling accuracy and the tractability brought by the block-fitting. We first illustrate how the block-correlation model can be adopted to generate random channel realizations within the FBL-FAS framework. Let $\boldsymbol{g}_u \in \mathbb{C}^{N}$ denote the channel coefficient vector of the $u$-th user, and let $\boldsymbol{\Sigma} \in \mathbb{C}^{N \times N}$ represent the corresponding spatial correlation matrix. Owing to the even placement of ports and the assumption of Clarke’s correlation model, the matrix $\boldsymbol{\Sigma}$ takes the form of a Toeplitz matrix:
\begin{equation}\label{eq.Clarke}
	\small
        \boldsymbol{\Sigma}=\begin{pmatrix}a(0)&a(1)&a(2)&...&a(N-1)\\a(-1)&a(0)&a(1)&...&a(N-2)\\\vdots&\ddots&&\vdots\\a(-N+1)&a(-N+2)&...&a(-1)&a(0)\end{pmatrix},
\end{equation}
where the generation function is $ a(n)=\operatorname{sinc}\left(\frac{2\pi nW}{N-1}\right)$.
With correlation matrix $\boldsymbol{\Sigma}$, one could generate $\boldsymbol{g}_u$ feasibly via eigenvalue-based construction \cite[Eq.~5]{FAS_tutorial}:
\begin{equation}\label{eq:channel_eigenvalue}
        \boldsymbol{g}_u=\boldsymbol{Q}\boldsymbol{\Lambda}^{\frac{1}{2}}\boldsymbol{g}_{0},
\end{equation}
where $\boldsymbol{Q}$ denotes the matrix of left eigenvectors and $\boldsymbol{\Lambda}$ is the diagonal matrix of corresponding eigenvalues obtained obtained from the eigenvalue decomposition of the correlation matrix $\boldsymbol{\Sigma} = \boldsymbol{Q}\boldsymbol{\Lambda}\boldsymbol{Q}^{\mathrm{H}}$, and $\boldsymbol{g}_{0} \in \mathbb{C}^{N}\sim \mathcal{CN}\left(\boldsymbol{0},\sigma^2\boldsymbol{I}\right)$. However, {\em directly analyzing the statistical distribution based on $\boldsymbol{\Sigma}$ is infeasible and overly complicated for any tractable analyses}.}
\par\indent \textcolor{blue}{Fortunately, the principal components of $\boldsymbol{\Sigma}$ can be captured by only a small number of dominant eigenvalues and their corresponding eigenvectors \cite[Fig.~3]{Block-Correlation}, \cite{FAS_tutorial}. Hence, the correlation matrix $\boldsymbol{\Sigma}$ can be approximated by a block-diagonal matrix:
\begin{equation}\label{eq:block_correlation_matrix}
        \widehat{\boldsymbol{\Sigma}}\in\mathbb{R}^{N\times N}=\begin{pmatrix}\mathbf{A}_1&\mathbf{0}&\cdots&\mathbf{0}\\\mathbf{0}&\mathbf{A}_2&\cdots&\mathbf{0}\\\vdots&&\ddots&\vdots\\\mathbf{0}&\mathbf{0}&\mathbf{0}&\mathbf{A}_B\end{pmatrix}.
\end{equation}
Each diagonal block $\boldsymbol{A}_b,~b \in \{1,2,\ldots,B\}$, is a constant correlation matrix of size $L_b \times L_b$, satisfying $\sum_{b=1}^{B}L_b = N$ (omit rounding errors). Intuitively, the correlation matrix characterizes the correlation among neighboring ports, where the correlation is typically strong due to the compact port spacing. The matrix $\widehat{\boldsymbol{\Sigma}}$ represents the fitted approximation of $\boldsymbol{\Sigma}$ according to \cite[Eq.~20]{Block-Correlation}, while offering a more convenient structure for analytical treatment. In particular, the block-correlation matrix exhibits the following common form:
\begin{equation} 
\mathbf{A}_b=\begin{pmatrix}
        1&\mu_b^2&\cdots&\mu_b^2\\
        \mu_b^2&1&\cdots&\mu_b^2\\
        \vdots&&\ddots&\vdots\\
        \mu_b^2&\cdots&\mu_b^2&1
\end{pmatrix},
\end{equation}
where $\mu_b^2\in \left(0.95,0.99\right),~b \in \{1,2,\ldots\}$ is the correlation constant. In summary, in this work, the Clarke's correlation model in \eqref{eq.Clarke} is adopted and the corresponding block diagonal approximation in \eqref{eq:block_correlation_matrix} is adopted for theoretical tractability.}

\section{Main Results}\label{sec.3}
This section introduces our main findings in terms of PDF derivations and fast calculation in Sec.~\ref{sec.III-a}, further simplified expressions for ease of computation in Sec.~\ref{sec.PDF_simplification} and BLER bound approximation in Sec.~\ref{sec.PDF_simplification}. The goal is to achieve simplified statistical results while maintaining high prediction accuracy.
\subsection{PDF of $|g_{\mathrm{FAS}}|^2$ and BLER under Block-Correlation Model}\label{sec.III-a}
\par The PDF of the random variable $
|g_{\mathrm{FAS}}|^2 = \max\left\{|g_{u,1}|^2, |g_{u,2}|^2, \ldots, |g_{u,N}|^2\right\}$ for FBL-FAS under the block-correlation channel model is given in \eqref{eq:pdf_channel_response}, where $f_{\chi_2^{\prime 2}}(x;\lambda)$ and $F_{\chi_2^{\prime 2}}(x;\lambda)$ denote the PDF and CDF of a non-central chi-square distribution with $2$ degrees of freedom and non-centrality parameter $\lambda$, respectively. We explain the derivations leading to \eqref{eq:pdf_channel_response} at the top of next page. Specifically, we derive the PDF from the CDF expression given next.
\par\indent\textbf{\em Remark}: {\em Readily converted from \cite[Eq.~43]{Block-Correlation}, \cite{FBL-FAS}, the CDF of random variable $|g_{\mathrm{FAS}}|^2=\max\left\{|g_{u,1}|^2,|g_{u,2}|^2,\ldots,|g_{u,N}|^2\right\}$ is:}
\begin{equation}\label{eq:remarkA1}
	\begin{aligned}
		&F_{|g_{\mathrm{FAS}}|^2}=P\left\{|g_{\mathrm{FAS}}|^2\le t\right\}\\
		&=\prod_{b=1}^B\int_0^\infty\frac{e^{-\frac{r_b}{2}}}{2}\times\left[1-Q_1\left(\sqrt{\frac{\mu^2r_b}{1-\mu^2}},\sqrt{\frac{t}{1-\mu^2}}\right)\right]^{L_b}\mathrm{d}r_b
	\end{aligned}.
\end{equation}
\begin{figure*}[t!]
	\normalsize
	\begin{equation}\label{eq:pdf_channel_response}
		\footnotesize
		\begin{aligned}
			&f_{|g_{\mathrm{FAS}}|^2}(t)\\
			&=\sum_{b=1}^B\left\{\int_0^\infty\frac{e^{-\frac{r_b}{2}}L_b}{2}\left[F_{\chi_2^{\prime2}}(\frac{t}{1-\mu^2};\frac{\mu^2r_b}{1-\mu^2})\right]^{L_b-1}\frac{f_{\chi_2^{\prime2}}(\frac{t}{1-\mu^2};\frac{\mu^2r_b}{1-\mu^2})}{1-\mu^2}\mathrm{d}r_b\right\}
			\times\prod_{j\neq b}^B\left\{\int_0^\infty\frac{e^{-\frac{r_j}{2}}}{2}\left[F_{\chi_2^{\prime2}}\left(\frac{t}{1-\mu^2};\frac{\mu^2r_j}{1-\mu^2}\right)\right]^{L_j}\mathrm{d}r_j\right\},
		\end{aligned}
	\end{equation}
	\hrulefill
\end{figure*}
\begin{figure}[t!]
	\centering
	\includegraphics[width=3.2 in]{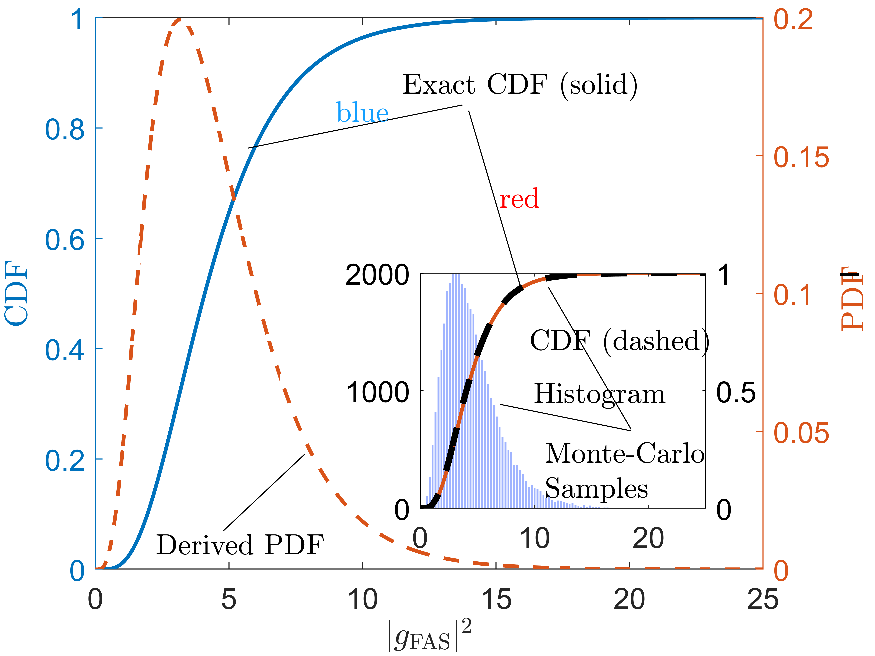}
	\caption{Illustration of block-correlation model's CDF and the PDF (block diagonal CDF in \eqref{eq:remarkA1}, the derived PDF in \eqref{eq:pdf_channel_response} and the corresponding Monte-Carlo samples.) with $N=10,~\mu^2=0.97,~W=0.5$.}
	\label{fig:PDF_CDF_Monte_Carlo}
\end{figure}
 For ease of descriptions, we denote \eqref{eq:remarkA1} by:
\begin{subequations}
	\begin{align}
		&F_{|g_{\mathrm{FAS}}|}=\prod_{b=1}^{B}G_b(t),\label{eq:r1-1}\\
		&G_b(t) = \int_0^\infty\frac{1}{2}e^{-r_b/2}\left[1-Q_1\left(a_b(r_b),\sqrt{a_t}\right)\right]^{L_b}\mathrm{d}r_b,\label{eq:r1-2}\\
		&a_b(r_b)=\sqrt{\frac{\mu^2}{1-\mu^2}r_b},~\sqrt{a_t}=\sqrt{\frac{t}{1-\mu^2}}.\label{eq:r1-3}
	\end{align}
\end{subequations}
It is clear that by differentiating CDF, one could obtain the corresponding PDF by $f_{|g_{\mathrm{FAS}}|^2}=\frac{\mathrm{d}F_{|g_{\mathrm{FAS}}|^2}}{\mathrm{d}t}$. However, to simplify the derivative procedures, the following identities are considered:
\begin{itemize}
	\item[-] The equivalence between Marcum-$Q$ and non-central chi-square:
	\begin{equation}\label{eq:M_chi}
		1-Q_1(a,\sqrt{x})=F_{\chi_2^{\prime2}}(x;a^2),\quad x\geq0,
	\end{equation}
	where $F_{\chi_2^{\prime2}}(x;a^2)$ is the non-central chi-square CDF with degrees-of-freedom 2 and non-centrality parameter $a^2$.
	\item[-] If $x$ in \eqref{eq:M_chi} is a function of variable $t$, i.e., $x(t)$, the following derivative can be established:
	\begin{equation}\label{eq:M_chi2}
		\frac{\mathrm{d}F_{\chi_2^{\prime2}}(x(t);a^2)}{\mathrm{d}t}=\frac{\mathrm{d}x(t)}{\mathrm{d}t}f_{\chi_2^{\prime2}}(x(t);a^2),
	\end{equation}
	where $f_{\chi_2^{\prime2}}(x;a)=\frac{1}{2}e^{-\frac{x+a}{2}}I_0(\sqrt{a x})$ is the corresponding chi-square distribution's PDF.
\end{itemize}
Thereby, if one consider indentity \eqref{eq:M_chi} in \eqref{eq:r1-1} and then use the derivative rule $f_{|g_{\mathrm{FAS}}|^2}(t)\equiv\frac{\mathrm{d}F_{|g_{\mathrm{FAS}}|^2}(t)}{\mathrm{d}t}=\sum_{b=1}^B\left(\frac{\mathrm{d}G_b(t)}{\mathrm{d}t}\right)\prod_{\begin{array}{c}j\neq b\end{array}}^BG_j(t)$ with identity in \eqref{eq:M_chi2}, \eqref{eq:pdf_channel_response} is obtained, which completes the proof.
\par\indent While \eqref{eq:pdf_channel_response} manifests complex integral structures, it can be feasibly computed by the Gauss--Laguerre quadrature method \cite{book1,book2} as follows:
\begin{itemize}
	\item[-] For an integrand of the form $e^{-x}f(x)$, the integral can be approximated as:
	\begin{equation}\label{eq.GL_A}
		\int_0^\infty e^{-x}f(x)\,\mathrm{d}x \;\approx\; \sum_{i=1}^{N_{\mathrm{GL}}} w_i f(x_i),
	\end{equation}
	where $x_i$ are the \emph{nodes} and $w_i$ are the corresponding \emph{weights}. The parameter $N_{\mathrm{GL}}$ is the expansion order, which determines the approximation accuracy (typically, choosing $N_{\mathrm{GL}}$ up to $64$ is sufficient but it can be well-tuned catering to different requirements).
	\item[-] According to the Golub--Welsch theorem \cite{book1,book2}, the values of $(x_i,w_i)$ can be derived by performing an eigenvalue decomposition on the following Jacobi matrix:
	\begin{equation}\label{eq:Jacobi}
		\boldsymbol{J}=\begin{pmatrix}
			\alpha_0 & \sqrt{\beta_1} & 0 & \cdots & 0 \\
			\sqrt{\beta_1} & \alpha_1 & \sqrt{\beta_2} & \cdots & 0 \\
			0 & \sqrt{\beta_2} & \alpha_2 & \cdots & 0 \\
			\vdots & \vdots & \ddots & \ddots & \sqrt{\beta_{N-1}} \\
			0 & 0 & \cdots & \sqrt{\beta_{N-1}} & \alpha_{N-1}
		\end{pmatrix},
	\end{equation}
	where $\alpha_n = 2n + 1$ and $\beta_n = n^2$.
	\item[-] The eigenvalue decomposition of \eqref{eq:Jacobi}, i.e., $\boldsymbol{J}=\boldsymbol{W}\boldsymbol{D}\boldsymbol{W}^{\mathrm{T}}$, provides the nodes $x_i$ as the $N$ largest eigenvalues (diagonal entries of $\boldsymbol{D}$). The corresponding weights $w_i$ are obtained from the squared elements of the first row of $\boldsymbol{W}$.
\end{itemize}
\par\indent For illustration of block-correlation modeling, the CDF in \eqref{eq:remarkA1}, the derived PDF in \eqref{eq:pdf_channel_response}, and the Monte Carlo samples (generated via block-correlation approximation) obtained from \eqref{eq:channel_eigenvalue} are plotted in Fig.~\ref{fig:PDF_CDF_Monte_Carlo}, where the channel model setups are generated with $N=10$, $\mu^2=0.97$, and $W=0.5$. Notably, consistent with \cite{Block-Correlation}, the CDF in \eqref{eq:remarkA1} assumes $\sigma^2=2$, which is adopted throughout the simulations in this work\footnote{It is straightforward to extend the CDF and PDF to different variances $\sigma^2$, which can be done by a standard change of variables in the PDF/CDF.}.
\textcolor{blue}{\subsection{Simplified PDF of $|g_{\mathrm{FAS}}|^2$ via Gamma Approximation}\label{sec.PDF_simplification}
As observed in Fig.~\ref{fig:PDF_CDF_Monte_Carlo}, for $t\ge 0$ the distribution of 
$f_{|g_{\mathrm{FAS}}|^2}(t)$ is {\em unimodal and right-skewed}, a key characteristic of the Gamma distribution family. This motivates the use of a Gamma approximation to simplify \eqref{eq:pdf_channel_response} and substantially reduce computational complexity. To approximate the random variable $t=|g_{\mathrm{FAS}}|^2$ by a Gamma distribution, we first compute its expectation and variance:
\begin{equation}\label{eq.E_V}
	\begin{aligned}
		\mathbb{E}\{t\}
		&= \int_0^{\infty} t\, f_{|g_{\mathrm{FAS}}|^2}(t)\,\mathrm{d}t,\\
		\mathrm{Var}\{t\}
		&= \mathbb{E}\{t^2\}-\left(\mathbb{E}\{t\}\right)^2\\
		&= \int_0^{\infty} t^2 f_{|g_{\mathrm{FAS}}|^2}(t)\,\mathrm{d}t
		-\left(\int_0^{\infty} t\, f_{|g_{\mathrm{FAS}}|^2}(t)\,\mathrm{d}t\right)^2.
	\end{aligned}
\end{equation}}
\par\indent \textcolor{blue}{We then assume $t \sim \mathcal{G}(k,\theta)$ with PDF and CDF:
\begin{equation}
	\begin{aligned}
		f_{\mathcal{G}}(t)
		&=\frac{1}{\Gamma(k)\theta^k}t^{k-1}e^{-t/\theta},\\[1mm]
		F_{\mathcal{G}}(t)
		&=\frac{\gamma\!\left(k,t/\theta\right)}{\Gamma(k)}, \quad t\ge 0,
	\end{aligned}
\end{equation}
where $\Gamma(\cdot)$ and $\gamma(\cdot,\cdot)$ denote the Gamma function and lower incomplete Gamma function, respectively. Matching the first two moments of the Gamma distribution with those in \eqref{eq.E_V} yields:
\begin{equation}\label{eq:Gamma_params}
	\begin{aligned}
		k\theta &= \mathbb{E}\{t\},\qquad 
		k\theta^2 = \mathrm{Var}\{t\},\\
		\Rightarrow\quad
		k &= \frac{\mathbb{E}\{t\}^2}{\mathrm{Var}\{t\}},\qquad
		\theta = \frac{\mathrm{Var}\{t\}}{\mathbb{E}\{t\}}.
	\end{aligned}
\end{equation}}
\par\indent \textcolor{blue}{With these parameters, the PDF and CDF of the channel strength in \eqref{eq:pdf_channel_response} and \eqref{eq:remarkA1} can be approximated as:
\begin{equation}\label{eq.GA_pdf_cdf}
	\begin{aligned}
		f_{|g_{\mathrm{FAS}}|^2}(t) \approx \frac{1}{\Gamma(k)\theta^k}t^{k-1}e^{-t/\theta},
		F_{|g_{\mathrm{FAS}}|^2}(t) \approx \frac{\gamma\!\left(k,t/\theta\right)}{\Gamma(k)},
	\end{aligned}
\end{equation}
where $k$ and $\theta$ are determined by \eqref{eq:Gamma_params}.}
\par
\begin{figure}[t!]
	\centering
	\includegraphics[width=3 in]{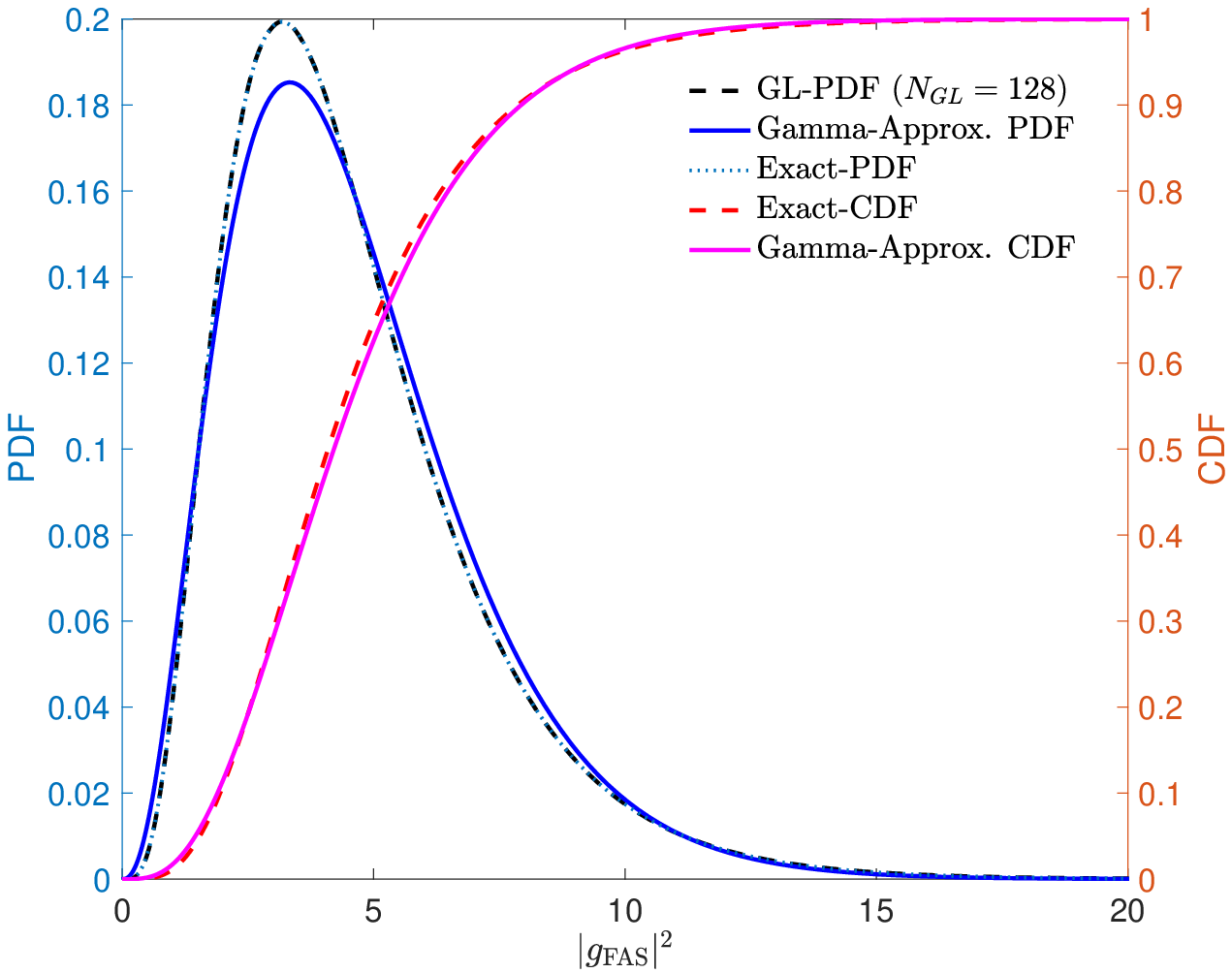}
	\caption{Illustration of block-correlation model's CDF and the PDF (exact CDF in \eqref{eq:remarkA1}, exact PDF in \eqref{eq:pdf_channel_response}, the Gauss-Laguerre-approximated PDF with $N_{\mathrm{GL}}=128$ by \eqref{eq.GL_A} and the Gamma-distribution-approximated PDF by \eqref{eq.GA_pdf_cdf}.) with $N=10,~\mu^2=0.97,~W=0.5$.}
	\label{fig:Gamma_Approximation_PDF_CDF}
\end{figure}

\indent \textcolor{blue}{To illustrate the accuracy of the proposed approximation, Fig.~\ref{fig:Gamma_Approximation_PDF_CDF} compares the exact CDF/PDF, the Gauss--Laguerre approximation with $N_{\mathrm{GL}}=128$ weights from \eqref{eq.GL_A}, and the Gamma-approximated PDF in \eqref{eq.GA_pdf_cdf}, for $N=10$, $\mu^2=0.97$, and $W=0.5$. The results confirm both the high fidelity of the Gauss--Laguerre approach and the effectiveness of the proposed Gamma-approximation method.}
\par\indent \textcolor{blue}{Finally, we remark on computational complexity. Computing the exact PDF via Gauss--Laguerre quadrature requires $\mathcal{O}(B\,N_{\mathrm{GL}})$ operations per evaluation point $t$. {\em In contrast, the proposed Gamma-approximated PDF is in closed form and requires only $\mathcal{O}(1)$ operations, with a negligible one-time cost for moment matching.} This makes the Gamma approximation highly attractive for large-scale or real-time evaluations.}
\subsection{BLER Approximation}\label{BLER_ApproBLER_Appro}
\par\indent\textcolor{blue}{\textbf{\em Remark}: \textit{For a given instantaneous SNR $\gamma$, the finite-blocklength BLER can be expressed as \cite{FBL1}:}
\begin{equation}\label{eq:bler_normal_approx}
	\epsilon(\gamma)
	=
	Q\!\left(
	\frac{C(\gamma) - R_c}{\sqrt{V_{\mathrm{dis}}(\gamma)/M}}
	\right),
\end{equation}
\textit{where $C(\gamma)=\frac{1}{2}\log_2(1+\gamma)$ is the channel capacity,
	$V_{\mathrm{dis}}(\gamma)=\frac{\gamma}{2}\frac{\gamma+2}{(\gamma+1)^2}\log_2^2(e)$ is the channel dispersion,
	$R_c=\frac{U}{M}$ is the code rate, and $Q(x)=\int_x^\infty\frac{1}{\sqrt{2\pi}}e^{-\tfrac{t^2}{2}}\mathrm{d}t$ is the standard $\mathrm{Q}$-function.}}
\par\indent \textcolor{blue}{We first rewrite the instantaneous SNR for fluid antennas. Since port selection yields the channel strength $t=|g_{\mathrm{FAS}}|^2$, the instantaneous SNR becomes $\gamma(t)=\frac{|g_{\mathrm{FAS}}|^2}{\sigma^2_{\eta}}$,
where the PDF and CDF of $t$ have been approximated in \eqref{eq.GA_pdf_cdf}.}
\par\indent \textcolor{blue}{To avoid the computational burden associated with the $Q$-function, we derive a high-SNR approximation of \eqref{eq:bler_normal_approx}. Using $\gamma\to\infty$ and the Gaussian tail approximation, we obtain:
\begin{equation}\label{eq.high_SNR_approx}
	\begin{aligned}
		V_{\mathrm{dis}}(\gamma)
		&= \frac{\gamma}{2}\frac{\gamma+2}{(\gamma+1)^2}\log_2^2(e)
		~\xrightarrow[\gamma\to\infty]{}~
		V_\infty\triangleq\frac{1}{2}\log_2^2(e), \\
		Q(x)
		&\approx
		\frac{1}{\sqrt{2\pi}\,x}
		\exp\!\left(-\frac{x^2}{2}\right),
		\qquad x\to+\infty.
	\end{aligned}
\end{equation}}
\par\indent \textcolor{blue}{Therefore, the high-SNR BLER approximation is:
\begin{align}\label{eq.BLER_approximation}
	\epsilon(\gamma)
	\approx
	\frac{1}{\sqrt{2\pi}\,x(\gamma)}\,
	\exp\!\left(-\frac{x^2(\gamma)}{2}\right),
\end{align}
where  $x(\gamma)
=
\sqrt{\frac{M}{V_\infty}}
\Biggl(\frac{1}{2}\log_2(1+\gamma)-R_c\Biggr)$ and $
V_\infty = \frac{1}{2}\log_2^2(e)$.
\textit{This simplified representation eliminates the integral in \eqref{eq:bler_normal_approx} and greatly reduces computational complexity.} Notably, approximation in \eqref{eq.high_SNR_approx} is expected to scale up the BLER and upperbounds the exact one.
\begin{figure}[t!]
	\centering
	\includegraphics[width=2.8 in]{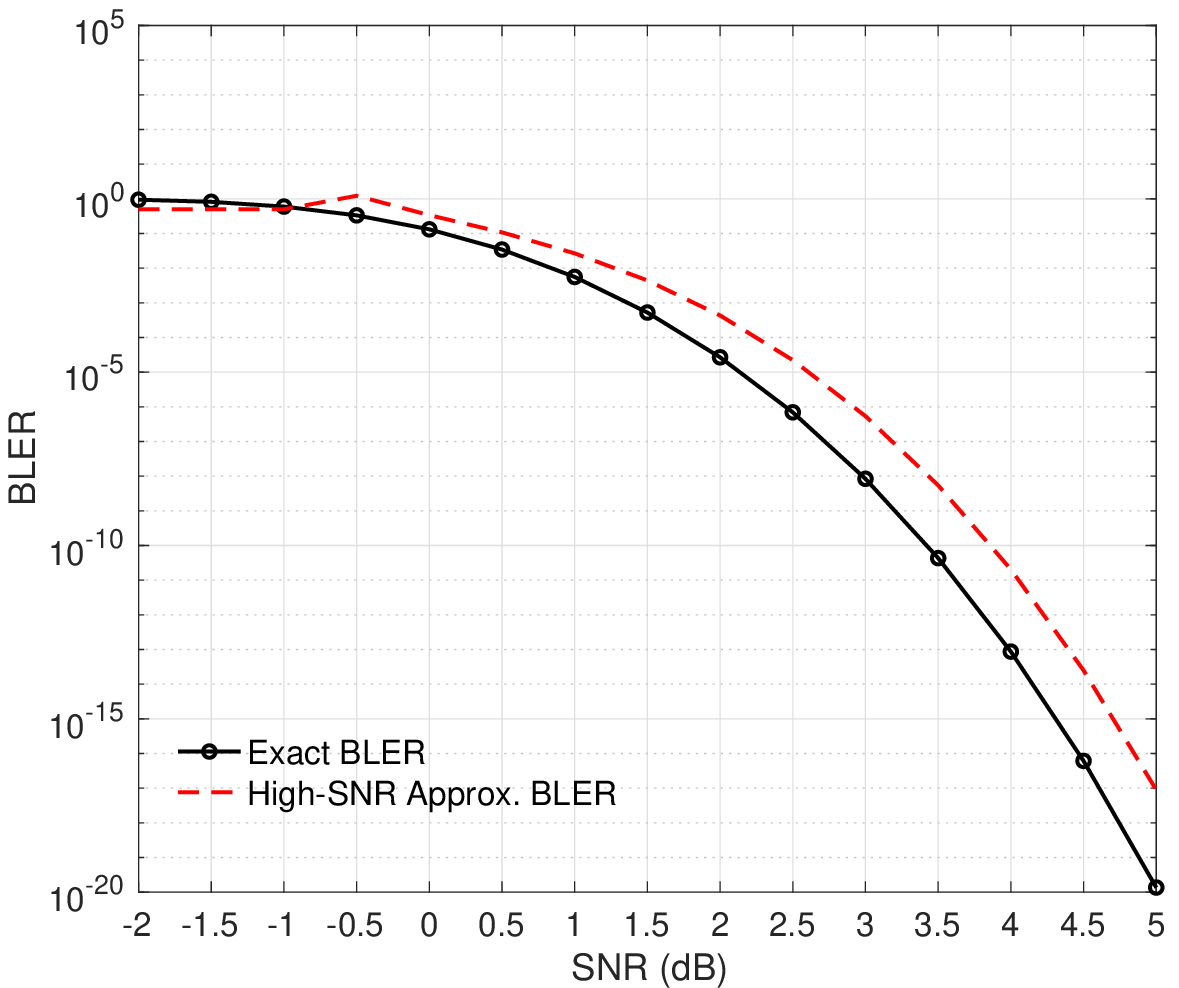}
	\caption{BLER comparison between exact and high-SNR approximation for $\sigma^2=2$, $|g_{\mathrm{FAS}}|^2=1$, $U=100$, and $M=400$.}
	\label{fig:BLER_Approximation}
\end{figure}}
\par\indent \textcolor{blue}{Fig.~\ref{fig:BLER_Approximation} compares the exact BLER from \eqref{eq:bler_normal_approx} and the approximation in \eqref{eq.BLER_approximation}. As expected, the high-SNR approximation is simplified but manifests higher BLER than the exact expression since $\frac{\gamma}{2}\frac{\gamma+2}{(\gamma+1)^2}\rightarrow 1$ only when $\gamma \rightarrow +\infty$.}
\par\indent \textcolor{blue}{Since both \eqref{eq:bler_normal_approx} and \eqref{eq.BLER_approximation} depend on the channel strength $|g_{\mathrm{FAS}}|^2$, whose PDFs (exact and approximated) are given in \eqref{eq:pdf_channel_response} and \eqref{eq.GA_pdf_cdf}, the average BLER can now be efficiently evaluated as:
\begin{equation}
	\bar{\epsilon}
	=
	\int_{0}^{+\infty}
	f_{|g_{\mathrm{FAS}}|^2}(t)\,
	\epsilon(t)
	\mathrm{d}t,
\end{equation}
where all required components have been significantly simplified, enabling fast and accurate BLER prediction for fluid-antenna-enabled systems.}
\section{Numerical Results}\label{sec.4}
\textcolor{blue}{In this section, simulations are conducted to demonstrate the advantages of FAS under the FBL assumption. The spatial diversity gain is verified by comparison with a conventional $L$-antenna system (fixed and independent antennas). The BLER of the conventional $L$-antenna system can be evaluated by substituting $\gamma=\frac{t_L}{\sigma^2_{\eta}}$ into \eqref{eq:bler_normal_approx}, where $t_L=\sum_{l=1}^{L}|g_l|^2$ denotes the channel strength after equalization (e.g., maximum-ratio combining/transmission) and $g_l, l\in\{1,\ldots,L\}$ are independent Rayleigh variables with zero mean and variance $\sigma^2=2$. Accordingly, $t_L\sim\mathcal{G}(L,\sigma^2)$ follows a Gamma distribution with PDF $f_{t_L}(t)=\frac{1}{\Gamma(L)\sigma^{2L}}t^{L-1}e^{-t/\sigma^2},~t\geq0$. The average BLER of the conventional $L$-antenna system is:
\begin{equation}
	\mathrm{Benchmark:}~\bar{\epsilon}_{L}=\int_{0}^{+\infty}\epsilon_{t_L}(t)f_{t_L}(t)\mathrm{d}t.
\end{equation}
{\em Notably, FBL-FAS activates only a single antenna (one RF chain), whereas a conventional system is equiped with $L$ RF chains whose advantage is conspicuous compared to FAS.} Unless otherwise specified, $U=100$ and $L=400$ are fixed, yielding a code rate of $\frac{100}{400}=0.25$ for a stringent short-packet scenario. The block-correlation constant is set as $\mu^2=0.97$.
\begin{figure}[t!]
	\centering
	\includegraphics[width=3 in]{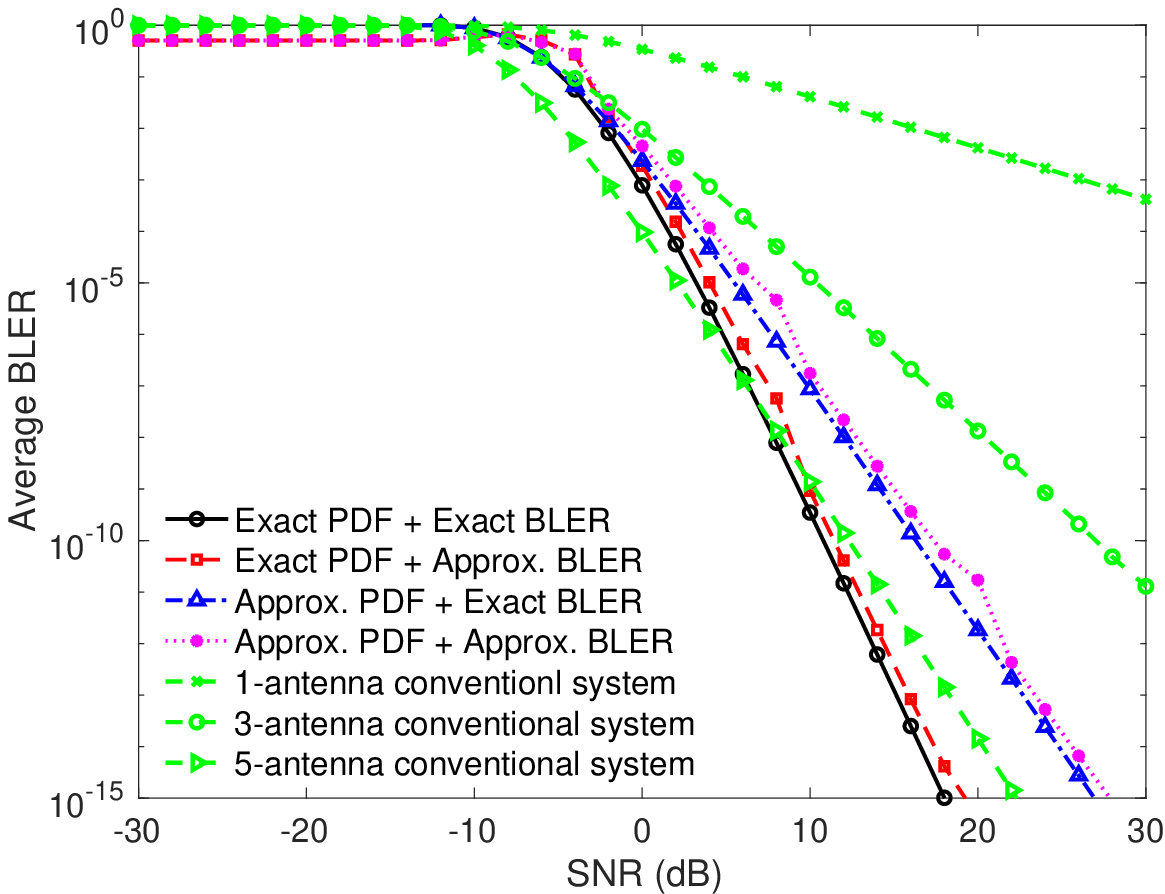}
	\caption{BLER versus SNR with $N=10$ available ports, $L\in\{1,3,5\}$, and $W=1$. All simulations are conducted within block-correlation model and the exact PDF refers to block-diagonal expression in \eqref{eq:pdf_channel_response} and the approximated PDF refers to the Gamma-approximated expression in \eqref{eq.GA_pdf_cdf}.}
	\label{fig:BLER_SNR_Benchmark}
\end{figure}}
\par\indent \textcolor{blue}{Fig.~\ref{fig:BLER_SNR_Benchmark} compares FBL-FAS and conventional systems with $L\in\{1,3,5\}$ when $N=10$ and $W=1$. The accuracy of both the channel-strength PDF approximation and the BLER expression is confirmed. Moreover, FBL-FAS exhibits significant random coding gain over the conventional multi-antenna system.
\begin{figure}[t!]
	\centering
	\includegraphics[width=3 in]{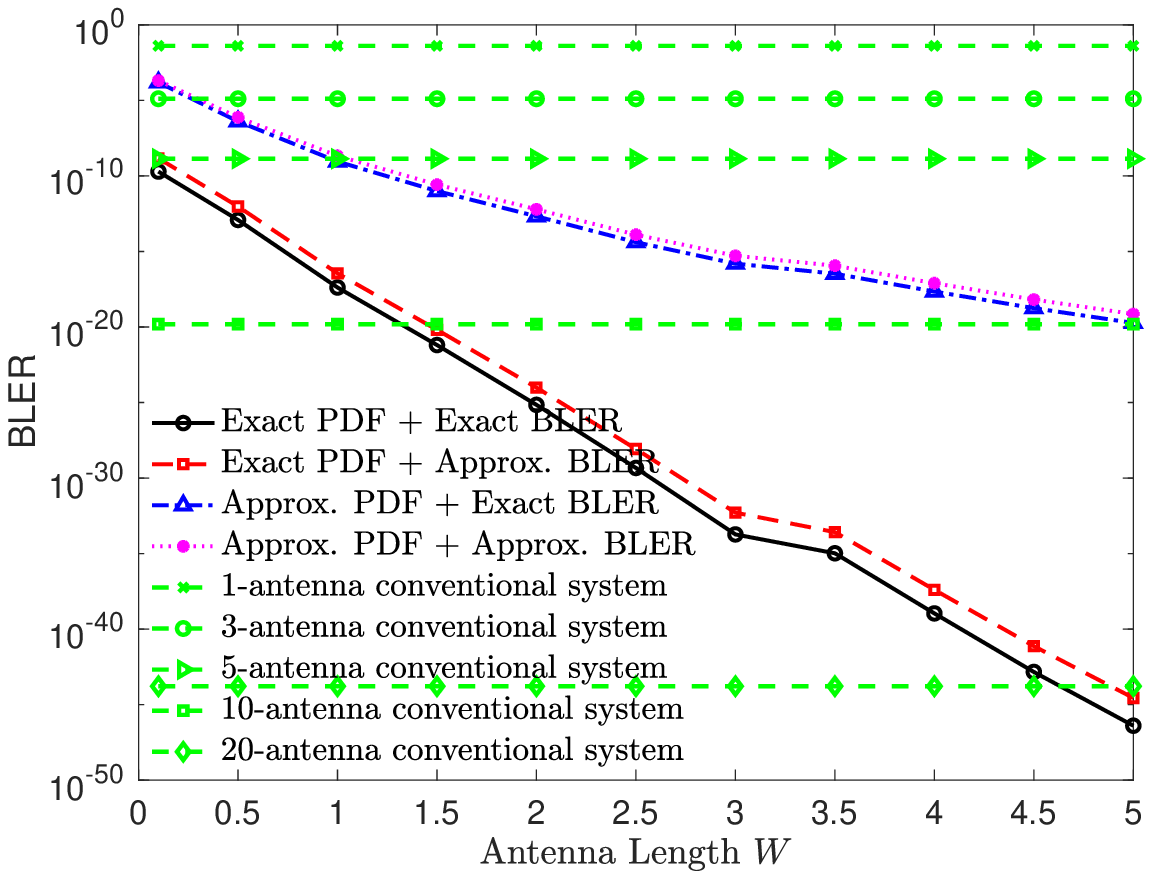}
	\caption{BLER versus aperture with $N=50$ available ports, $L\in\{1,3,5,10,20\}$, and SNR $=10$ dB.}
	\label{fig:BLER_W_Benchmark}
\end{figure}}
\par\indent \textcolor{blue}{Fig.~\ref{fig:BLER_W_Benchmark} shows the BLER of FBL-FAS versus aperture with $N=50$ and SNR $=10$ dB. For $W=5$, the minimum spacing between ports is $0.1$ (normalized) under uniform placement. Notably, with $W=5$, the FBL-FAS achieves performance comparable to a conventional system with $20$ antennas. With increasing aperture and fixed $N$, the available spatial diversity grows and the BLER decreases monotonically.}
\par\indent \textcolor{blue}{It is also observed that the PDF approximation error increases when comparing Fig.~\ref{fig:BLER_SNR_Benchmark} and Fig.~\ref{fig:BLER_W_Benchmark}. When $N=10$, the gap between the exact and approximate PDFs is minor, whereas it becomes larger for $N=50$. This is expected since the Gamma approximation in \eqref{eq.GA_pdf_cdf} is implicitely correlated to $N$ and the envolop of Gamma distribution is relatively regimented while \eqref{eq:pdf_channel_response} becomes increasingly unimodal with larger $N$. In practice, $N$ is constrained by the limit of hardware design, and moderate configurations (e.g., Fig.~\ref{fig:BLER_SNR_Benchmark}) provide accurate performance prediction with reduced complexity.}
\section{Conclusion}\label{sec.5}
\textcolor{blue}{This work investigates the random coding performance limit of FBL communication with fluid antennas. Both the channel-strength model and a tractable BLER expression are developed to enable efficient performance prediction. We derive the PDF of block-correlated FAS and present an accurate approximation. In addition, the BLER expression is simplified to an integral-free form, significantly reducing computational complexity. Numerical results confirm that the spatial diversity provided by fluid antennas yields substantial gains for short-packet communication systems, highlighting a promising design paradigm.}

\end{document}